\documentclass[aps,floatfix,preprint,nofootinbib,amsmath]{revtex4}
\usepackage{epsfig}
\begin{document}
\def\mod{{\rm mod}}
\def\tr{{\rm tr}}
\def\rhob{{\boldsymbol\rho}}
\def\sigmab{{\boldsymbol\sigma}}
\def\be{\begin{equation}}
\def\ee{\end{equation}}
\def\Do{{\Delta^{(3)}_o}}
\def\De{{\Delta^{(3)}_e}}
\def\lb{\langle}
\def\rb{\rangle}
\title{
 Odd-even staggering in nuclear binding and the liquid-drop model.
}
\author{
W.A.~Friedman}
\affiliation{
Department of Physics and Institute of Nuclear Theory,
Box 351560\\
University of Washington
Seattle, WA 98915 USA\\
}
\begin{abstract}

The trends with mass number are examined for the odd-even-staggering (OES)
in nuclear binding of neutrons and protons through the conventional measures
 $\Delta^{(3)}$. 
The large differences previously observed between these trends for
even and odd values of
these measures is found to arise, in part, from the slow variation
of binding energies  with mass and charge which provides a
background contribution. This background
is estimated with the liquid-drop
model, and accounts for the greater difference found in proton removal
relative to neutron removal. The differences which persist after
backgrounds are removed can not be treated in the conventional
liquid-drop model but require the addition of a new term. Such a term
is investigated, and its effect on specific values of the OES is calculated.
The liquid-drop fitting is also applied to a set of separation energies
constrained to match the specific set of nuclei used to  determine
the observed values for the odd  $\Delta^{(3)}$.
The resulting
fit for the pairing term is compared to the average value of even and odd
measures.
The effect on this value  of the new liquid-drop term is observed,
and the change in background when the
new term is included is also used as an
alternate method for determining the difference between trends of the even and odd
values of the OES.

\end{abstract}
\maketitle
\section{Introduction}
The abrupt changes in binding energy 
as one goes from
a nucleus with an even number of neutrons (or protons) to its neighbor with
a odd number of such nucleons is know as odd-even-stagger (OES). This effect  has been much
studied and is often associated with BCS pairing. However it is known
to involve other effects. \cite{BF1}. 

The experimental observations of OES  have been 
quantitatively examined  by several measures. 
One  of these is $\Delta^{(3)}$, which, for  
changing neutron number $N$ at fixed $Z$, is defined in terms of 
binding energies $BE$ by
\begin{equation}
\Delta^{(3)}(N)= \pm .5(BE(N+1)+BE(N-1)-2BE(N))
\end{equation}
An analogous definition $\Delta^{(3)}(Z)$ applies to changing proton number
with fixed $N$.
The  plus sign is used when  $N$ is odd, and minus when $N$ 
is even. This convention guarantees that  
all measures are positive when nuclei with even neutron numbers
are more bound than those with odd. 

The observed trends with mass number $A$ show a marked difference
between cases with  $N$ odd,  $\Delta^{(3)}_o(N)$,  and those with
$N$ even, $\Delta^{(3)}_e(N)$.
The fits to an $A$-dependence for neutrons, reported in ref.1,  were found
to be
$(.82+24./A)$MeV and $(.94+41./A)$MeV respectively.
For protons the fits are found to have an even 
greater difference, $(.75+25./A)$Mev and $(1.11+59./A)$Mev
respectively. If the  staggering arose solely
from BCS pairing one would expect no difference between the  behavior
for even and odd cases.
The purpose of this paper is to further examine this difference making
use of the liquid-drop model.
We treat here the situation for both neutron and proton removal.
    
As seen in Eq. 1  the  measures $\Delta^{(3)}$ are simply proportional
to the second difference with respect to neutron number ( or proton number).
In addition to the OES, 
there are also  slower mass trends involving  variations of binding energy with neutron and
proton number. These variations can  also contribute to the 
second differences.
Here, we call the contribution from this type of variation "background".  
The OES  contribution is positive
for both even and odd cases of $\Delta^{(3)}$.  The background
contribution, however,  changes sign due to the convention 
indicated in the definitions. However there would be
little difference in its magnitude in going from one nucleus to its
neighbor if the background variation 
is slow enough. 

 In order to distinguish the specific OES contributions from background effects we label these  
by the short-hand notations $ne$ and $no$ for neutrons and $pe$ and $po$ for protons.
For the respective background contribution 
we use the symbol  $\beta$
such that
\begin{eqnarray}
\Delta^{(3)}_o(N)=no+\beta_n \\
\Delta^{(3)}_e(N)=ne-\beta_n \nonumber \\ 
\Delta^{(3)}_o(P)=po+\beta_p \nonumber \\ 
\Delta^{(3)}_e(P)=pe-\beta_p \nonumber
\end{eqnarray}

When forming the average of the even and odd measures, $.5(\Delta^{(3)}_e+\Delta^{(3)}_o)$, the contributions
from the background $\beta$ 
cancel. Therefore this average is affected only by
the contributions from OES.  On the other hand, in forming
the difference of the measures, e.g., $ .5(\Delta^{(3)}_e-\Delta^{(3)})$, the result is strongly influenced by the background $\beta$.

\section{Background}
We  use here the liquid-drop model to assist in exploring the differences 
between the
even and odd measures. This model is particularly suited for
describing the slow variation leading to the background. 

The standard liquid-drop model, or more simply the
Weizsacker mass formula, was devised to deal in a simple
phenomenological way with physical features of nuclear binding energy that are common
to classical  systems as well. These include a bulk interaction
which is treated by a volume term proportional to
the number of nucleons $A$. In addition,
surface tension and  Coulomb interactions are treated by two additional
terms proportional to $A^{2/3}$ and $Z^2/A^{1/3}$ respectively.
To account for some of the quantum and isospin effects
peculiar to nuclei, an asymmetry term is also required.
This is taken proportional to $(N-Z)^2/A$.
Another feature of  nuclear binding energies
is the OES.
To a limited extent this feature has, until now, been included in the liquid-drop
mass formula  through a pairing term taken proportional
to $0.5((-1)^N+(-1)^Z)$ multiplied by function of mass number $\mu(A)$.
While the
pairing term of the standard mass  model
can indeed incorporate the effects
of BCS pairing, it cannot incorporate some of the features
arising from other physics. In particular 
it cannot treat the physics leading to the difference
in the mass trends
between $\Delta^{(3)}_o$ and
$\Delta^{(3)}_e $.                                   
This requires a further
enhancement of the liquid-drop mass formula.
By studying the observe difference in the mass trends of experimental
data for odd and even values of $\Delta^{(3)}$
we can also learn about the nature  of such an extension of the liquid-drop approach.

Because the liquid-drop model provides simple analytic expressions for each of the 
four slowly
contributions, it is convenient to evaluate the
second partial derivatives of these with respect to either neutron number $N$
or charge $Z$ . These can be immediately associated
with the second differences of 
$\Delta^{(3)}$ which contribute to the backgrounds $\beta_n$ and $\beta_p$. 

The function $\mu(A)$ in the pairing term  
is frequently assumed to have the form $c/A^{1/2}$.
This can be replaced by the form $\mu(A) = c_1+c_2/A$ with 
little affect on the other terms
and no effect on the rms deviation of the overall 
fit to the table of measured masses. As
pointed out in ref.1, the later form is more transparent 
for indicating the physical origins of this term. 

\section{New extension of liquid-drop model}

The liquid-drop model as described above is
unable to provide a difference between the OES contributions associated
with  even and odd cases, $ne$ and $no$, and $pe$ and $po$,
since $\mu$ is assumed to
vary little from one nucleus to its neighbor.
In order to
be able to  obtain a difference, in the context of the 
liquid-drop model, an  
interesting new effect must be added.

In ref. 1  we examined possible physical sources which lead to 
such differences. This was done from the point of
view of mean-field models. 
To summarizes,  we found that a source for these differences arose from the degeneracy
of the time reversed orbits.  This degeneracy introduces a non-smooth
progression of single particle energies with the successive filling
of  nucleon orbits.
That is, the single-particle energies change only after a  pair of orbits
are filled.  The interaction terms which reflect the wave-functions of these
orbits show a similar lack of smoothness with progressive filling.
The expected BCS pairing, on the other hand,
does not provide a difference.

For convenience we define as $\delta_n$ and $\delta_p$  the expected
non-zero difference between even and odd OES, so that
\begin{eqnarray}
\delta_n=ne-no \\
\delta_p=pe-po \nonumber
\end{eqnarray}

To see how $\delta$ can be incorporated into the liquid-drop treatment
we next examine the neutron case as a specific  example. 
We add a new smoothly varying part $\eta$ to the liquid-drop formula.  This would contributed
to the background of $ne$ and $no$ with a magnitude $.5d^2\eta_n/dA^2$
but with opposite signs.

To link the $\eta_n$ to $ne$ and $no$ 
we consider the following expressions for each,
involving two contributions to the OES. 
\begin{eqnarray}
ne \equiv no+\delta=\mu-.5d^2\eta/dA^2  \\
no=\mu +.5d^2\eta/dA^2 \nonumber
\end{eqnarray}
These expressions  reflect the additional background from $\eta$,
and $\mu$ is the average of $ne$ and $no$.
Taking the difference we have 
\begin{equation}
\delta=-d^2\eta/dA^2
\end{equation}
This gives a direct link between $\eta$ and $\delta$, and shows
that a non-zero value of $\delta$ 
requires
a non-zero function  $\eta(A)$.

\section{ Data}

We turn next to the measured values of $\Delta^{(3)}$.
If the full background $\beta$  consists of
a part $\beta_0$ from the conventional liquid-drop and the additional
new term  $.5d^2\eta/dA^2$,
we obtain
\begin{eqnarray}
\Delta^{(3)}_e(N)=ne-{\beta_0}_n-.5d^2\eta_n/dA^2  \\
\Delta^{(3)}_o(N)=no+{\beta_0}_n+.5d^2\eta_n/dA^2 \nonumber \\ 
\Delta^{(3)}_e(P)=ne-{\beta_0}_p-.5d^2\eta_p/dA^2 \nonumber \\ 
\Delta^{(3)}_o(N)=no+{\beta_0}_p+.5d^2\eta_p/dA^2 \nonumber
\end{eqnarray}

From Eq.2  and the link between $\eta_n$ and $\delta_n$, Eq.5,  we obtain the difference
\begin{equation}
.5(\Delta^{(3)}_e(N)-\Delta^{(3)}_o(N))=-{\beta_0}_n + \delta_n
\end{equation}
(Note that had $\eta_n(A)$ been missing, the $\delta_n$ on the right hand side of Eq. 7
would have been multiplied by a factor of $0.5$.)
Similar expressions follow for protons.

The experimental mass trends for
the left hand side of Eq.7  have been
presented in ref.1. These trends 
have been fit by
($c_1 +c_2/A$) and the explicit values $c_1$ and $c_2$ have been
reported, having the values of $(0.06+8./A)$MeV for neutrons
and $(0.18 + 17./A)$MeV for protons.  

We estimate here the mass trends for  
$\beta_0$ using the following procedure. 
We fit coefficients for the standard 5-term liquid-drop model to
the complete table of measured nuclear binding energies \cite{mass}.  
For this
procedure we used the on-line program \cite{Scid}.
The results for the fitting coefficients are 15.7, 17.6. 23.4 and 0.71
for volume, surface, asymmetry, and Coulomb terms. These are 
similar to those commonly reported in text books \cite{text}.

The volume term, proportional to $A$, has a vanishing second derivative 
so to calculate $\beta_0$ we need only the remaining three.
With the best fit coefficients, the second partial derivatives with respect to neutron
number of the respective $A$ and $Z$
functions are easily found.  
The values ${\beta_0}_n$ and ${\beta_0}_p$ were  evaluated for the
explicit set of $N$ and $Z$  
used to obtain the measured $\Delta^{(3)}_o(N)$ and $\Delta^{(3)}_o(P)$. These 
sets of values were then fit by
a two-term expression $c_1+c_2/A$
giving the resulting trends for $\beta_0$  
\begin{eqnarray}
{\beta_0}_n= (-0.0066 - 16.43/A)  MeV \\
{\beta_0}_p= (-0.087 - 32.79/A) MeV \nonumber
\end{eqnarray}

From Eq.7 we then estimate the mass trends for $\delta_n$ and $\delta_p$
using the coefficients provided in table 3 of ref.1 for the observed difference of 
measures,
\begin{eqnarray}
\delta_n=(0.05-8./A) MeV \\
\delta_p=(0.09-16./A)MeV \nonumber
\end{eqnarray}
The full background $\beta_0-.5\delta$ then gives
\begin{eqnarray}
\beta_n=(-0.04-12./A) MeV\\
\beta_p=(-0.14- 25./A) MeV \nonumber
\end{eqnarray}
The greater values for the proton background result in the greater
difference between the observed even and odd measures.

Using this dependence for $\beta$ and the values of $\delta$ we deduce
\begin{eqnarray}
  \\
no=(0.82+24./A)+(0.04+12./A)=(0.86+36./A)\nonumber \\
ne=(0.86+36./A) + (0.05-8./A)=(0.91+28./A)\nonumber \\ 
po=(0.75+25./A)+(0.14+25./A)=(0.89 + 50./A) \nonumber \\ 
pe= (0.89 + 50./A)+(0.09-16./A)=(0.98+34./A) \nonumber
\end{eqnarray}
Finally, as a numerical check, the average of $ne$ and $no$  and the average of
$pe$ and $po$ give the same results as the average of the respective
$\Delta^{(3)}_e$ and $\Delta^{(3)}_o$ as they must, by construction.

\section { Separation energies for constrained sets of nuclei}
The measure $\Delta^{(3)}$ is also proportional to the difference
between neighboring separation energies. Therefore we studied fits
to the sets of measured neutron and proton
separation energies.
For this we also used the liquid-drop approach but fit the
coefficients of the various terms to the sets  of respective  separation
energies (neutron and proton) rather than the binding energies.  In this study we have
included only nuclei which are
actually involved in obtaining the sets of values for  $\Delta^{(3)}_o(N)$ 
and $\Delta^{(3)}_o(P)$ reported in 
ref.1.
The constraints on the data are outlined there and, corresponding to these,
for neutrons we have used separation energies for nuclei with $Z$ even, 
$N>Z$, and $Z$ greater or equal
to 8 and $N$ greater or  equal to 11. For protons we have used  $N$ even, $N>Z$, $N$ great
or equal to 12 and $Z$ greater or equal to 9.
Because of the close link between OES and the separation energies
this promised to provide additional confirmation of the
approach to backgrounds taken above with the fit to binding energies.

With the constrained  set of nuclei the rms deviations  for the 
best fit to the separation energies, with the 5-term conventional
form of the liquid-drop formula, was 0.62 MeV for neutrons and .57 MeV
for protons. (Note, this is the rms from the best fit to the
separation energies and not binding energies.)  The coefficients of the
pairing term was $12.4/A^{1/2}$MeV for neutrons and $13.9/A^{1/2}$MeV for protons. 
When the form of the pairing
term was taken  as $(c_1+c_2/A)$  the best fit was given
by $(0.897+29.7/A)$MeV for neutrons and $(.96+37.2/A)$MeV for protons. These fits provide
the same rms deviations. In Section III we 
found the average value of the odd and even measures for neutrons
and protons to be 
was $(0.88+33./A$)MeV and $(0.93+42./A)$ MeV as reported found in Ref.1.
This is in relatively close agreement with this direct fit to the sets of
separation  energies using only the
four slowly varying terms of the liquid-drop.

Since the 5-term fit lacked the
flexibility to provide a difference
between $no$ and $ne$, for example,  it could be 
expected that the rms deviations might be improved
by repairing this deficiency with the additional term $\eta$ discussed
above.
One of the contributions to $\delta$ is anticipated to be proportional
to the level spacing of the single-particle energies. A good fit to the
mass trends of this  quantity is associated with the form  $1/A$.
Consequently one might anticipate $\eta(A)$ to have the form
$ A ln(A)$
in order that its second derivative might provide $\delta$.

When a additional term 
of the form  ($c Aln(A)$) 
was added to the 5-term liquid-drop formula, the
rms deviation value was indeed reduced to .58 MeV for neutrons and .54 MeV
for protons. 
The values of the best fit coefficients for the additional terms
are found to be 
$c=-9.98$  and $c=+18.75$ for neutrons.
and protons respectively. 
However the coefficients of the 4 traditional terms
change from their original  values when the extra $\eta $ terms were added.

We next considered the values of $\delta_n$ and $\delta_p$ which might be
obtained from $d^2\eta/dA^2$. If the added terms ($cAln(A)$) did not affect the other
coefficients then the fitting coefficients themselves would lead directly
to the form of $\delta$. However this is not the case, and the values
of $\beta_0$ are perturbed in the new fitting process. To incorporate
this change we consider the "effective" value of $.5d^2/dA^2$ to be the
difference between the background for the augmented liquid-drop (with
the $\eta$ added) the traditional liquid-drop (without $\eta$).
We thus consider
\begin{equation}
.5d^2\eta/dA^2=\Delta \beta_0 +0.5c/A
\end{equation}

The net change $\Delta \beta_0$ 
when $\eta$ is added is $(-0.038+9.72/A)$MeV and $(-.04-4.73/A)$MeV
 for neutrons
and protons respectively, and the constants $c$ 
are coefficients of  $Aln(A)$
from the fits.
The value of $\delta$ is then given by  multiply $.5d^2\eta/dA^2$ by -2.

This procedures produces  values
for $\delta$ of $(0.08-9.5/A)$MeV and $(0.08-9.3/A)$MeV respectively
for neurons and protons. 
This is qualitatively similar to the differences $\delta$
$(0.05-8./A)$MeV and $(0.09-16./A)$MeV which 
were  estimated in Eq.11 of Section IV.
The $1/A$ terms in the expressions for $\delta$ appear to have  negative
coefficients. If the physics behind this term is, as suggested in Ref. 1, 
the sum  of a single-particle level
spacing term (providing a positive coefficient of order 45) and potential energy term (providing
a negative coefficient), then the potential energy term seems to  dominate but
be of similar magnitude as the level spacing term.

The addition of the $Aln(A)$ also leads to direct fits for the pairing function
which are slightly closer to the observed average of $\Delta^{(3)}_e$
and $\Delta^{(3)}_o$ reported 
reported in Ref.1.
For neutrons the fit provided $(.88+33./A)$MeV, precisely the observed value.
For protons it provided $(.96+38./A)$MeV, slightly closer to observed
value than without the additional term.

\section{ Liquid-drop Pairing term fits}
We also examined the sensitivity to the pairing term when the 
liquid-drop formula
is applied to the set of 
separation energies values used in forming $\Delta^{(3)}_o(N)$. For
this purpose we performed a series of 
least-square fits starting with only the pairing term ( and none
of the slowly varying terms)
and then added, one-by-one, the other four terms of the model.
When all 5-terms of the liquid-drop are used the fit to the
pairing term provides a coefficient of about
$12.4/A^{1/2}$ quoted above. When no terms except the pairing term 
is used the fit gives
a value about $13./A^{1/2}$ with a huge rms deviation. As more and more
of the terms are  added the fits to the pairing term 
trend slowly downward toward the fit obtained
with the full complement of terms,  and the rms 
deviations decrease substantially.
What is remarkable is the small range over which the best fit
pairing coefficient changes.
This  suggests that the pairing term is strongly determined by the set
of separation energies with little influence from the other important
terms. Those terms are, of course, quite important for obtaining a reasonable fit.
In Table I we demonstrate this by providing the pairing coefficients
and the rms deviations for fits involving increasing numbers of 
liquid-drop terms.

\begin{table}[h1]
\caption{Fits of $c$ for the coefficient of the pairing term $c/A^{1/2}$
with increasing number of terms in liquid-drop
formula. The fit is to neutron separation energies for even $Z$, $N>Z$, $Z$  greater
or equal to 8, and for two values of $N_{min}$, 10 and 11}
\begin{tabular}{|c|c|c|c|}
\colrule
Terms used& $N_{min}=10$&$N_{min}=11$& rms \\
&&& MeV\\
\colrule
pair. & 12.9 & 13.1 & 8.02 \\
pair.+vol.& 12.7 & 12.7 & 1.81 \\
pair.+vol.+surf.& 12.7 & 12.7 & 1.81 \\
pair.+vol.+surf.+coul. & 12.6 &12.5 & 1.39 \\
pair,+vol.+surf.+coul.+asym& 12.4 &12.4 &0.62  \\
\colrule
\end{tabular}
\end{table}

We next examined the correspondence between fits in which the pairing coefficients has a
2-term form ($c_1+c_2/A$) and those in which it has  the 1-term form
($c/A^{1/2}$). 
We considered the  set of  2-term pairing 
coefficients ($c_1+c_2/A)$ for which the values of $A$ were taken for each
nucleus belonging to the 
set of nuclei used for forming
the $\Delta^{(3)}_o(N)$. We then fit this set of values
with 
a curve employing the 1-term form $c/A^{1/2}$.
The results of this exercise for values of $c$ giving the best fit
are show in Table II.

\begin{table}[h2]
\caption{Fit by the curve $c/A^{1/2}$  to values  of $c_1 + c_2/A$ for each of nuclei associated
with $\Delta^{(3)}_o(N)$. The values shown are
$c$ for each combination of $c_1$  and $c_2$}
\begin{tabular}{|c|c|c|c|}
\colrule
value of $c_1$ & $c_2$=29. & $c_2$=30 & $c_2$=31 \\
\colrule
0.89 & 12.24 & 12.35 & 12.46 \\
0.90 & 12.34 & 12.45 & 12.56 \\
\colrule
\end{tabular}
\end{table}

This seems to explain why similar fits result with the
use of either the 2-term or  1-term forms. 
It also suggests
that a change of 0.1 in the value of the coefficient of the
1-term fit corresponds roughly to a change of 1. in the value
or $c_2$ or .01 in the value of $ c_1$.

\section{Summary and conclusion}
The mass trends of OES have been examined using the measures
 $\Delta^{(3)}$. The observed binding energies indicate a large  difference
been even and odd values of these quantities. It is  pointed
out that the slow variations of the binding energies, with mass and
charge, can significantly influence this difference by contributing
background contributions $\beta$ in addition to the OES. We have studied
this influence by using the results of the liquid-drop model to
describe these variations. While the backgrounds are significant
for all the measures, they are found to be greater for the proton removal
than for neutron removal. Consequently this
can account to the greater differences in the observed values. This effect
probably arises from the greater influence of the Coulomb energy and
the asymmetry energy with a predominance of neutron number
over proton number in the mass tables.

After the
background is removed
differences in OES are still anticipated from effects discussed in ref. 1,
and such differences $\delta$ cannot be treated by the
conventional terms employed
in the liquid-drop model. Consequently we propose adding to the
liquid-model a new contribution
arising from the specific physics contributing
to this difference between even and odd trends in the staggering.
We have examined the new contributing function, labeled here as $\eta(A)$,
and find that $d^2\eta/dA^2=-\delta$. We have combined the observed difference
$.5(\Delta^{(3)}_e-\Delta^{(3)}_o)$ and the liquid-drop background to
estimate $\delta$ for neutrons and protons. These estimates provide
$(0.05-8./A)$MeV for neutrons and $(.09-16./A)$Mev for protons.
For comparison, the
difference between $\Delta^{(3)}_e$ and $\Delta^{(3)}_o$ provides
the much greater values of
$(0.12+16./A)$MeV for neutrons and $(0.36+34./A)$MeV for protons.
Note the sign change in the coefficient to the $1/A$ terms.

We have also applied the liquid-drop model fit directly to the
separation energies for the specific set of nuclei used to constructed
the observed $\Delta^{(3)}_o$. This procedure, without the additional
contribution $\eta$
led to a pairing function close to the observed averages
of the odd and even OES mass trends. We have also added a contribution
$\eta$, assuming a form $Aln(A)$ motivated by the expectation that
$\delta$, and hence $d^2\eta/dA^2$, might follow a form $1/A$.  This addition
improved
the rms fit and also brought the pairing term function closer
to the observed average of even and odd OES; it providing an exact match
for neutron removal and  qualitative match for protons. Adding the
$\eta$ term also led to a change of the
background contribution which could be  associated with an effective  value of
$d^2\eta/dA^2$, and
consequently $\delta$.  This method of determining $\delta$ provided
values in qualitative agreement with those obtained
by removing an estimated background:
$(0.08-9./A)$ compared to (0.05-8./A) for neutrons, and $(0.08-9./A)$
compared to $(0.09-16./A)$ for protons.

 A study of the liquid-drop fit to the observed separation energies
suggests that the data strongly determines the liquid-drop pairing term values,
and that these values are little influenced by the presence
or absence of the other terms in the model. Namely, the best fit for the
pairing term changes from
$13.0/A$ to $12.4/A$ as the number or other liquid-drop
terms ranges from zero to the
conventional four.  We also made a detailed comparison
of the two forms for the pairing function,
$c/A$ and $(c_1+c_2/A)$, where a change of 0.1 in $c$ corresponds to
a change of 0.01 in $c_1$ or 1.0 in $c_2$.

\end{document}